\documentclass[11pt]{article}

\usepackage{epsfig}

\begin{document}
%%%%%%%%%%%%%%%%%%%%%%%%%%%%
\title{{\bf Optimization of thermal noise in multi-loop pendulum
suspensions for use in interferometric gravitational-wave 
detectors}}
\author{Constantin Brif \\
{\small\it LIGO Project, California Institute of Technology,
Pasadena, CA 91125}}
\date{}
\maketitle

\begin{abstract}
We study the thermal-noise spectrum of multi-loop pendulum 
suspensions for test masses in interferometric gravitational-wave 
detectors. The dependence of the thermal noise on suspension 
parameters and on properties of the wire material is discussed for 
the situation in which the losses are dominated by the internal 
friction in the pendulum wires.
\vskip 2mm

\end{abstract}

\section{Introduction}

The thermal noise is expected to be one of the main limitations
on the sensitivity of long-baseline interferometric gravitational-wave 
detectors like LIGO and VIRGO \cite{Abram92,Caron97}.
Thermal fluctuations of internal modes of the interferometer's test 
masses and of suspension modes will dominate the noise spectrum at 
the important frequency range between 50 and 200 Hz (seismic noise 
and photon shot noise dominate for lower and higher frequencies, 
respectively). The thermal fluctuations in pendulum suspensions
were studied both theoretically and experimentally in a number of
works (see, e.g.\ Refs.~\cite{Saul90,GoSa94,Logan93,GiRa93,GiRa94,%
Gao95,Brag94a,Brag94b,Brag96,Cagn96,Rowan97a,DaKa97,HuSa98}).
The predictions of the thermal-noise spectrum in interferometric 
gravitational-wave detectors combine theoretical models (with the 
fluctuation-dissipation theorem of statistical mechanics 
\cite{Callen} serving as a basis) and experimental measurements of 
quality factors of systems and materials involved. It is usually
assumed that losses in the suspensions will occur mainly due to the
internal friction in the wires, which is related to anelasticity 
effects \cite{Zen48,NoBe72}. This assumption will be correct only
provided that all the losses due to interactions with the external 
world (friction in the residual gas, dumping by eddy currents, 
recoil losses into the seismic isolation system, friction in the 
suspension clamps, etc.) are made insignificant by careful 
experimental design.

In the present work we consider a multi-loop pendulum suspension
and study the dependence of the thermal-noise spectrum on 
properties of the wire material and on suspension parameters.
The thermal-noise spectral density $x^2 (\omega)$ depends strongly 
on the type of the internal friction in the wires. We consider two 
possibilities: (i) the wire internal friction with a constant loss 
function and (ii) the thermoelastic damping mechanism \cite{Zener}. 
The main conclusion is that the thermal noise can be reduced by 
increasing the number of suspension wires, especially in the case 
of the thermoelastic damping. This conclusion is valid as long as 
the dissipation due to the friction in the suspension clamps is 
insignificant.

\section{Thermal-noise spectrum for a pendulum suspension}

In interferometric gravitational-wave detectors, the test masses 
are suspended as pendulums by one or two loops of thin wires. We 
will consider a multi-loop suspension with the wires attached to 
the bob near the horizontal plane which cuts the bob through its 
center of mass. We will also assume that the mass of the wires is 
much smaller than the mass of the bob. In such a multi-loop 
suspension the rocking motion of the test mass is essentially
suppressed and the main contribution to the thermal-noise 
spectrum is due to the pendulum mode and the violin modes.
Then one can write the suspension thermal-noise spectral density 
as a sum,
\begin{equation}
x^2 (\omega) = x^2_{{\rm p}} (\omega) + x^2_{{\rm v}} (\omega) ,
\end{equation}
of the pendulum-mode contribution, $x^2_{{\rm p}} (\omega)$, and
of the violin-modes contribution, $x^2_{{\rm v}} (\omega)$.

According to the fluctuation-dissipation theorem, the 
pendulum-mode contribution can be expressed as \cite{Saul90}
\begin{equation}
\label{eq:x2p-1}
x^2_{{\rm p}} (\omega) = \frac{ 4 k_B {\mathcal T} }{\omega M}
\frac{\omega_{{\rm p}}^2 \phi_{{\rm p}} (\omega)}{
( \omega_{{\rm p}}^2 - \omega^2 )^2 
+ \omega_{{\rm p}}^4 \phi_{{\rm p}}^2 } ,
\end{equation}
where $k_B$ is Boltzmann's constant, ${\mathcal T}$ is the 
temperature, $M$ is the pendulum mass, $\phi_{{\rm p}} (\omega)$ is 
the loss function, $\omega_{{\rm p}} = (g/L)^{1/2}$ is the pendulum 
frequency, $g$ is the acceleration due to the Earth gravity field, 
and $L$ is the pendulum length. 
Note that the spectral density $x^2 (\omega)$ is written explicitly 
in terms of the angular frequency $\omega$, but in fact the density
is with respect to the linear frequency $f = \omega/2\pi$ and 
$x^2 (\omega)$ is measured in units of m$^2$/Hz.

The loss function $\phi$ is a measure of the energy dissipation. 
Let ${\mathcal E}$ be the total energy of a dissipative oscillator 
(assuming that the losses are small) and $\Delta {\mathcal E}$ 
be the energy dissipated per cycle. Then 
\begin{equation}
\phi = \frac{ \Delta {\mathcal E} }{2 \pi {\mathcal E}} .
\end{equation}
The energy of the pendulum consists of two parts: the gravitational 
energy ${\mathcal E}_{{\rm gr}}$ and the elastic energy 
${\mathcal E}_{{\rm el}}$ due to the bending of the wire. The
gravitational energy is lossless; provided that all the losses due 
to interactions with the external world are made insignificant by 
careful experimental design, the assumption is made that the losses 
are dominated by internal friction in the wire material. 
Consequently, $\Delta {\mathcal E} = \Delta {\mathcal E}_{{\rm el}}$,
and one obtains 
\begin{equation}
\phi_{{\rm p}} = \xi_{{\rm p}} \phi_{{\rm w}} ,
\end{equation}
where 
$\phi_{{\rm w}} = \Delta {\mathcal E}_{{\rm el}} /(2 \pi 
{\mathcal E}_{{\rm el}} )$ 
is the loss function for the wire itself which occurs due to anelastic 
effects in the wire material, and 
$\xi_{{\rm p}} = ( {\mathcal E}_{{\rm el}} / 
{\mathcal E}_{{\rm gr}} )_{{\rm p}}$ 
is the ratio between the elastic energy and the gravitational energy 
for the pendulum mode. The elastic energy depends on how many wires 
are used and how they are attached to the pendulum bob. In the
multi-loop configuration we consider, the wires bend both at the top 
and the bottom, so $\xi_{{\rm p}} \simeq (k_e L)^{-1}$, where
$k_e^{-1} \simeq (E I/T)^{1/2}$ is the characteristic distance scale
over which the bending occurs. Here, $T$ is the tension force in the 
wire, $E$ is the Young modulus of the wire material, and $I$ is the 
moment of inertia of the wire cross section  
($I = \frac{1}{2} \pi r^4$ for a cylindrical wire of radius $r$).
For a suspension with $N$ wires (the number of wires is twice the
number of loops), $T = M g/N$, and one obtains
\begin{equation}
\xi_{{\rm p}} \simeq \frac{ N \sqrt{T E I} }{M g L}  =
\frac{1}{L} \sqrt{ \frac{E I N}{M g} } .
\end{equation}
For LIGO suspensions, $f_{{\rm p}} = \omega_{{\rm p}}/2 \pi$ is 
about 1 Hz. This is much below the working frequency range  
(near 100 Hz), so we may assume $\omega_{{\rm p}} /\omega \ll 1$.
Also, the loss function is very small, $\phi_{{\rm p}} < 10^{-5}$.
Then the pendulum-mode contribution to the thermal noise spectrum 
is
\begin{equation}
\label{eq:x2p-2}
x^2_{{\rm p}} (\omega) \simeq \frac{ 4 k_B {\mathcal T} 
\omega_{{\rm p}}^2 \phi_{{\rm p}} (\omega)}{M \omega^5}
= \frac{ 4 k_B {\mathcal T} }{ L^2 } \sqrt{ \frac{g E I N}{M^3} } 
\frac{\phi_{{\rm w}} (\omega)}{\omega^5} .
\end{equation}

The contribution of the violin modes to the thermal noise spectrum 
is given by
\begin{equation}
\label{eq:x2v-1}
x^2_{{\rm v}} (\omega) = \frac{4 k_B {\mathcal T}}{\omega} 
\sum_{n=1}^{\infty} \frac{ \mu_n^{-1} \omega_n^2 \phi_n (\omega) 
}{ ( \omega_n^2 - \omega^2 )^2 + \omega_n^4 \phi_n^2 } ,
\end{equation}
where $n = 1,2,3,\ldots$ is the mode number. 
The angular frequency of the $n$th mode is
\begin{equation}
\omega_n = \frac{n \pi}{L} \sqrt{\frac{T}{\rho}} \left[
1 + \frac{2}{k_e L} + \frac{1}{2} \left( \frac{n \pi}{k_e L}
\right)^2 \right] ,
\end{equation}
where $\rho$ is the linear mass density of the wire. For heavily 
loaded thin wires like in LIGO, $k_e^{-1} \ll L$, so
\begin{equation}
\label{eq:om_n}
\omega_n \simeq \frac{n \pi}{L} \sqrt{\frac{T}{\rho}} .
\end{equation}
This is just the angular frequency of the $n$th transverse 
vibrational mode of an ideal spring. 
The effective mass of the $n$th violin mode is
\begin{equation}
\mu_n = \frac{1}{2} N M \left( \frac{\omega_n}{\omega_{{\rm p}}} 
\right)^2 
\simeq \frac{ \pi^2 M^2 }{2 \rho L} n^2 , 
\end{equation}
where we took expression (\ref{eq:om_n}) for $\omega_n$ and 
$T = M g/N$. This effective mass arises because the violin 
vibrations of the wire cause only a tiny recoil of the test 
mass $M$.
The loss function for the $n$th violin mode is
\begin{equation}
\phi_n = \xi_n \phi_{{\rm w}} ,
\end{equation}
where $\xi_n = ( {\mathcal E}_{{\rm el}} / 
{\mathcal E}_{{\rm gr}} )_n$ is the ratio between the elastic energy 
and the gravitational energy. This ratio is \cite{GoSa94}
\begin{equation}
\label{eq:xi_n}
\xi_n = \frac{2}{k_e L} \left( 1 + \frac{ n^2 \pi^2 }{2 k_e L}
\right)  .
\end{equation}
Since $k_e L \gg 1$, for first several modes the energy ratio is 
approximately
\begin{equation}
\xi_n \simeq \xi_{{\rm v}} = \frac{2}{L} \sqrt{\frac{E I N}{M g}} .
\end{equation}
This expression takes into account only the contribution to the 
elastic energy due to wire bending near the top and the bottom. 
For higher violin modes, one should also consider the 
contribution due to wire bending along its length, which leads to 
Eq.~(\ref{eq:xi_n}).

Typical values of $f_1 = \omega_1 /2\pi$ are from 250 to 500 Hz.
If we are interested in the thermal spectral density near 
100 Hz, we can assume $\omega^2 \ll \omega_n^2$. Then we have 
approximately
\begin{equation}
\label{eq:x2v-2}
x^2_{{\rm v}} (\omega) \simeq \frac{ 8 k_B {\mathcal T} 
\omega_{{\rm p}}^2 }{ N M \omega} \sum_{n=1}^{\infty} 
\frac{ \phi_n (\omega) }{ \omega_n^4 } \simeq
\frac{ 8 k_B {\mathcal T} N \rho^2 L^3 }{ \pi^4 g M^3 \omega}
\sum_{n=1}^{\infty} \frac{ \phi_n (\omega) }{ n^4 } .
\end{equation}
One can see that the contributions of higher violin modes are very 
small due to the factor $n^{-4}$ in the sum. Taking 
$\phi_n = \xi_n \phi_{{\rm w}}$ and assuming $k_e L \gg 1$, we find 
the following expression for the violin-mode contribution to the 
thermal-noise spectrum,
\begin{equation}
\label{eq:x2v-3}
x^2_{{\rm v}} (\omega) \simeq 
\frac{8}{45} k_B {\mathcal T} \rho^2 L^2  
\sqrt{ \frac{ E I N^3 }{ g^3 M^7 } } 
\frac{ \phi_{{\rm w}} (\omega) }{\omega} .
\end{equation}

\section{Dependence of thermal noise on wire material and suspension 
parameters}

It can be seen from Eqs.~(\ref{eq:x2p-2}) and (\ref{eq:x2v-3}) that
the thermal noise increases with the area $A$ of the wire cross 
section. Therefore, it is desirable to use wires as thin as 
possible. However, the wire thickness may not be too small since 
the stress $\sigma = T/A$ in the wire may not exceed the breaking 
stress $\sigma_{{\rm br}}$. In fact, the wires are always operated
at a fixed fraction of their breaking stress, 
\begin{equation}
\sigma = \sigma_0 = \kappa \sigma_{{\rm br}} ,
\end{equation}
where $\kappa$ is a numerical coefficient. Typical values of 
$\kappa$ are from 0.3 to 0.5 (it is undesirable to have larger
values of $\kappa$ because then events of spontaneous stress 
release will contribute excess noise \cite{Ageev97}).
Thus for a given type of the wire material, the cross-section
area $A$ should be proportional to the pendulum mass $M$,
according to the relation $\sigma_0 = M g/(N A)$. 
For a cylindrical wire, one has $I = A^2 /2\pi$. Then we obtain
\begin{equation}
\label{eq:x2p-si}
x^2_{{\rm p}} (\omega) = \frac{4 k_B {\mathcal T}}{L^2}
\left( \frac{g^3 E}{2 \pi M N \sigma_0^2 } \right)^{1/2}
\frac{\phi_{{\rm w}} }{\omega^5} ,
\end{equation}
\begin{equation}
\label{eq:x2v-si}
x^2_{{\rm v}} (\omega) = \frac{8}{45} k_B {\mathcal T} 
\rho_v^2 L^2 
\left( \frac{g^3 E}{2 \pi M N^3 \sigma_0^6 } \right)^{1/2}
\frac{\phi_{{\rm w}} }{\omega} ,
\end{equation}
where $\rho_v = \rho/A$ is the volume mass density of the wire
which depends only on the material used.

All the parameters in Eqs.~(\ref{eq:x2p-si}) and (\ref{eq:x2v-si}) 
are easily measured except for the wire loss function 
$\phi_{{\rm w}}$. A number of experiments were recently performed 
\cite{GiRa94,DaKa97,HuSa98,KoSa93,Saul94,Rowan97b,Cagn99} to study 
internal losses of various wire materials (e.g., steel, tungsten, 
fused quartz, and some others). However, the exact 
frequency dependence of the wire loss function 
$\phi_{{\rm w}} (\omega)$ is not yet completely understood. In many 
experiments $\phi_{{\rm w}}$ was measured only at few frequencies 
and experimental uncertainty of results was often quite large. 
Moreover, there are discrepancies between results of different 
experiments. Therefore, it is sometimes difficult to make certain 
conclusions about the behavior of $\phi_{{\rm w}} (\omega)$. 

A well known dissipation mechanism for thin samples in flexure is 
the so-called thermoelastic damping \cite{Zener}. As a wire bends,
one side contracts and heats and the other expands and cools.
The resulting thermal diffusion leads to the dissipation of energy.
The corresponding loss function is
\begin{equation}
\label{eq:ted}
\phi_{{\rm w}}(\omega) = \Delta \frac{ \omega \bar{\tau} }{ 
1 + \omega^2 \bar{\tau}^2 } ,
\end{equation}
where $\Delta$ is the relaxation strength and $\bar{\tau}$ is
the relaxation time. The loss function has its maximum 
$\phi = \Delta/2$ at $\omega = \bar{\tau}^{-1}$ (this is called 
the Debye peak). This behavior is characteristic for processes
in which the relaxation of stress and strain is exponential and
occurs via a diffusion mechanism. For the thermoelastic damping,
one has \cite{Zener}
\begin{equation}
\Delta = \frac{E {\mathcal T} \alpha^2}{C_v} , \hspace{12mm}
\bar{\tau} \simeq \frac{d^2}{D} ,
\end{equation}
where $\alpha$ is the linear thermal expansion coefficient, $C_v$ is 
the specific heat per unit volume, $d$ is the characteristic distance 
heat must flow, and $D$ is the thermal diffusion coefficient, 
$D = \varrho/C_v$, where $\varrho$ is the thermal conductivity. For a 
cylindrical wire of diameter $d$, the frequency of the Debye peak is
\begin{equation}
\bar{f} = \frac{1}{2 \pi \bar{\tau}} \simeq 2.6 \frac{D}{d^2} .
\end{equation}
For thin metallic wires ($d \sim 100\ \mu$m) at the room temperature,
the Debye peak frequency is typically from few hundred Hz to few 
kHz. Therefore at the frequency range near 100 Hz, we are usually far 
below the Debye peak, and 
\begin{equation}
\label{eq:ted-low}
\phi_{{\rm w}}(\omega) \simeq \Delta \omega \bar{\tau}
= \beta A \omega ,
\end{equation}
where $\beta \simeq \Delta/(1.3 \pi^2 D)$. 

According to a recent experiment by Huang and Saulson \cite{HuSa98}, 
internal losses in stainless steel wires are in good agreement
with predictions of thermoelastic damping, with 
$\phi_{{\rm w}}(\omega)$ exhibiting the characteristic frequency
dependence of Eq.~(\ref{eq:ted}). On the other hand, the loss function 
for tungsten wires was nearly constant, increasing slightly at high
frequencies (above 500 Hz). $\phi_{{\rm w}}$ for tungsten wires 
increased with the wire cross-section area $A$, but the exact 
functional dependence of $\phi_{{\rm w}}$ on $A$ is unclear as only
three different wire diameters were examined. In some other
experiments, the loss functions for various materials were found to 
be nearly constant over a wide frequency range. In a recent 
experiment by Cagnoli et al.\ \cite{Cagn99}, internal damping
of a variety of metallic wires was found to be well modelled
by the loss function of the form
\begin{equation}
\phi_{{\rm w}}(\omega) = \phi_0 + \phi_{{\rm ted}}(\omega) ,
\end{equation}
where $\phi_{{\rm ted}}(\omega)$ is the thermoelastic-damping loss 
function of Eq.~(\ref{eq:ted}) and $\phi_0$ is a 
frequency-independent term. Unfortunately, the dependence of
$\phi_0$ on the wire diameter was not examined.
It can be assumed that the thermoelastic damping is a basic 
dissipation mechanism, but for some materials it is masked by other 
processes. When those additional losses (whose nature is still a 
matter of controversy) are small, the characteristic frequency 
dependence of Eq.~(\ref{eq:ted}) may be observed. However, when the 
losses due to the thermoelastic damping are very small (which happens, 
for example, in the case of thin Invar and tungsten wires), then 
additional losses prevail, leading to $\phi_{{\rm w}}$ which is 
nearly constant far from the Debye peak.

In what follows we will consider two possibilities: (i) a constant
loss function $\phi_{{\rm w}}$ and (ii) the loss function of 
Eq.~(\ref{eq:ted-low}) which is characteristic for the 
thermoelastic damping at frequencies well below the Debye peak. 
We might assume that for some materials the true behavior is 
somewhere between these two extreme variants. 
For example, for tungsten wires, $\phi_{{\rm w}}$ is nearly 
frequency-independent from 50 to 500 Hz, but still increases to 
some extent with the wire cross-section area $A$, as one should 
expect from Eq.~(\ref{eq:ted-low}).

\subsection{A constant loss function}

For a constant $\phi_{{\rm w}}$, the dependence of the 
thermal-noise spectrum on various parameters is given directly by 
Eqs.~(\ref{eq:x2p-si}) and (\ref{eq:x2v-si}).
For the pendulum-mode contribution, we find
\begin{description}
\item for constant $M$ and $\sigma_0$, 
$x^2_{{\rm p}} \propto N^{-1/2}$;
\item for constant $M$ and $N$, 
$x^2_{{\rm p}} \propto \sigma_0^{-1}$;
\item for constant $N$ and $\sigma_0$, 
$x^2_{{\rm p}} \propto M^{-1/2}$.
\end{description}
For the violin-modes contribution, we find
\begin{description}
\item for constant $M$ and $\sigma_0$, 
$x^2_{{\rm v}} \propto N^{-3/2}$;
\item for constant $M$ and $N$, 
$x^2_{{\rm v}} \propto \sigma_0^{-3}$;
\item for constant $N$ and $\sigma_0$, 
$x^2_{{\rm v}} \propto M^{-1/2}$.
\end{description}
The allowed stress $\sigma_0$ is a property of the wire material
(which is also true for $E$, $\rho_v$, and $\phi_{{\rm w}}$),
so changing $\sigma_0$ means taking wires made of different
materials. Clearly, it is desirable to have a material with
a large value of $\sigma_0$, but what decides is the value
of the factor 
$\Lambda_{{\rm w}} = E^{1/2} \phi_{{\rm w}}/\sigma_0$ 
for the pendulum mode and 
$\Lambda_{{\rm w}} = \rho_v^2 E^{1/2} \phi_{{\rm w}}/\sigma_0^{3}$ 
for the violin modes. The factor $\Lambda_{{\rm w}}$ comprises all
the parameters in $x^2$ which characterize the wire material.

%%%%%%%%%%%%%%%%%%%%%%%%%%%%%%%%%%%%%%%%%%%%%%%%%%%%%%%%%%%
\begin{figure}[htbp]
\label{fig:tung1}
\epsfxsize=0.72\textwidth
\centerline{\epsffile{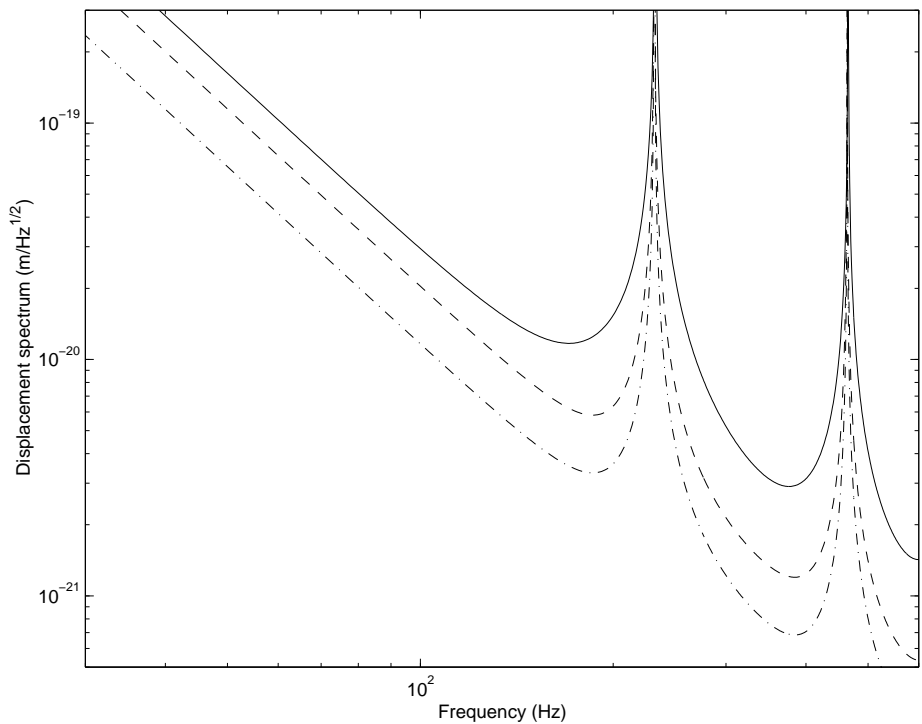}}
\caption{The thermal-noise displacement spectrum 
$\sqrt{x^2 (\omega)}$ for a multi-loop pendulum suspension with
tungsten wires: 
$N = 4$, $Q_{{\rm w}} = 1.3 \times 10^3$ (solid line);
$N = 16$, $Q_{{\rm w}} = 1.3 \times 10^3$ (dashed line);
$N = 16$, $Q_{{\rm w}} = 4.0 \times 10^3$ (dash-dot line).}
\end{figure}
%%%%%%%%%%%%%%%%%%%%%%%%%%%%%%%%%%%%%%%%%%%%%%%%%%%%%%%%%%%

One may see that taking multi-loop suspensions with large numbers 
of wires may help to reduce the thermal noise. As an example, let 
us consider tungsten wires of the type examined by Huang and 
Saulson \cite{HuSa98}. The relevant parameters are 
$E \simeq 3.4 \times 10^{11}$ Pa,
$\sigma_{{\rm br}} \simeq 1671$ MPa,
$\rho_v \simeq 1.93 \times 10^4$ kg/m$^3$.
We also take $M = 10.8$ kg, $L = 45$ cm and $\kappa = 0.5$ 
(the wires are operated at one half of their breaking stress),
like in suspensions of the LIGO test masses. 
According to the data by Huang and Saulson \cite{HuSa98}, the 
loss function is nearly frequency-independent from 50 to 500 Hz, 
but depends on the wire diameter.
For a two-loop suspension ($N = 4$), the wire diameter should be 
$d \simeq 200$ $\mu$m, and the corresponding quality factor
$Q_{{\rm w}} = \phi_{{\rm w}}^{-1}$ can be estimated to be 
$Q_{{\rm w}} \simeq 1.3 \times 10^3$. For an eight-loop suspension 
($N = 16$), the wire diameter should be $d \simeq 100$ $\mu$m, and 
the corresponding quality factor can be estimated to be 
$Q_{{\rm w}} \simeq 4.0 \times 10^3$. In Fig.~1 we 
plot the thermal-noise displacement spectrum $\sqrt{x^2 (\omega)}$ 
for the room temperature (${\mathcal T} = 295$ K) for three 
possibilities:
(a) $N = 4$, $Q_{{\rm w}} = 1.3 \times 10^3$; 
(b) $N = 16$, $Q_{{\rm w}} = 1.3 \times 10^3$;
(b) $N = 16$, $Q_{{\rm w}} = 4.0 \times 10^3$.
We see that for a constant loss function, the thermal noise is
reduced by increasing the number of wires. The spectral density
$x^2(\omega)$ scales as $N^{-1/2}$ for frequencies near 100 Hz
(where the pendulum mode dominates), in accordance with our
analysis. Also, if the decrease of $\phi_{{\rm w}}$ with the
wire diameter is taken into account, the increase in the number
of wires is even more helpful.

\subsection{Thermoelastic loss function}

If we take the loss function of Eq.~(\ref{eq:ted-low}), then
the thermal-noise spectrum is given by
\begin{equation}
\label{eq:x2p-ted}
x^2_{{\rm p}} (\omega) = \frac{4 k_B {\mathcal T}}{L^2} \beta
\left( \frac{g^5 E M}{2 \pi N^3 \sigma_0^4 } \right)^{1/2}
\frac{1}{\omega^4} ,
\end{equation}
\begin{equation}
\label{eq:x2v-ted}
x^2_{{\rm v}} (\omega) = \frac{8}{45} k_B {\mathcal T} 
\beta \rho_v^2 L^2 
\left( \frac{g^5 E M}{2 \pi N^5 \sigma_0^8 } \right)^{1/2} .
\end{equation}
The dependence of the thermal-noise spectrum on various parameters
can be characterized as follows. For the pendulum-mode contribution, 
we find
\begin{description}
\item for constant $M$ and $\sigma_0$, 
$x^2_{{\rm p}} \propto N^{-3/2}$;
\item for constant $M$ and $N$, 
$x^2_{{\rm p}} \propto \sigma_0^{-2}$;
\item for constant $N$ and $\sigma_0$, 
$x^2_{{\rm p}} \propto M^{1/2}$.
\end{description}
For the violin-modes contribution, we find
\begin{description}
\item for constant $M$ and $\sigma_0$, 
$x^2_{{\rm v}} \propto N^{-5/2}$;
\item for constant $M$ and $N$, 
$x^2_{{\rm v}} \propto \sigma_0^{-4}$;
\item for constant $N$ and $\sigma_0$, 
$x^2_{{\rm v}} \propto M^{1/2}$.
\end{description}
Now, the dependence of $x^2$ on the wire material is given by the
factor $\Lambda_{{\rm w}} = \beta E^{1/2} /\sigma_0^{2}$ 
for the pendulum mode and 
$\Lambda_{{\rm w}} = \beta \rho_v^2 E^{1/2} /\sigma_0^{4}$ 
for the violin modes. So, the value of the allowed stress
$\sigma_0$ in this situation is more important than for the case
of constant $\phi_{{\rm w}}$. 

One may see that in the case of the thermoelastic damping the 
thermal noise may be reduced to a larger extent by increasing the 
number of wires, as compared to the case of constant 
$\phi_{{\rm w}}$. As an example, let us consider wires made of 
stainless steel (AISI 302), which were examined by Huang and 
Saulson \cite{HuSa98}. The relevant parameters are 
$E \simeq 1.9 \times 10^{11}$ Pa,
$\sigma_{{\rm br}} \simeq 1342$ MPa,
$\rho_v \simeq 8.0 \times 10^3$ kg/m$^3$.
The losses are dominated by the thermoelastic damping mechanism. 
Taking
$\alpha \simeq 1.6 \times 10^{-5}$ 1/K, 
$C_v \simeq 4.8 \times 10^6$ J/(K m$^3$),
$\varrho \simeq 16.3$ J/(K m s) and 
${\mathcal T} = 295$ K, one obtains
$\Delta \simeq 3.0 \times 10^{-3}$ and 
$\beta \simeq 68.6$ s/m$^2$. 
We also take $M = 10.8$ kg, $L = 45$ cm and $\kappa = 0.5$,
like in suspensions of the LIGO test masses. 
The thermal-noise displacement spectrum $\sqrt{x^2 (\omega)}$ 
is plotted in Fig.~2 for three possibilities:
(a) $N = 4$ (then $d \simeq 224$ $\mu$m and 
$\bar{f} \simeq 176$ Hz); 
(b) $N = 8$ (then $d \simeq 159$ $\mu$m, and 
$\bar{f} \simeq 352$ Hz);
(b) $N = 16$ (then $d \simeq 112$ $\mu$m, and 
$\bar{f} \simeq 703$ Hz).
The conclusion is that the thermal noise may be significantly
reduced by increasing the number of wires. The numerical 
results confirm that the proportionalities
$x^2_{{\rm p}} \propto N^{-3/2}$ and
$x^2_{{\rm v}} \propto N^{-5/2}$ are valid for frequencies 
well below the Debye peak $\bar{f}$. 

%%%%%%%%%%%%%%%%%%%%%%%%%%%%%%%%%%%%%%%%%%%%%%%%%%%%%%%%%%%
\begin{figure}[htbp]
\label{fig:ss1}
\epsfxsize=0.72\textwidth
\centerline{\epsffile{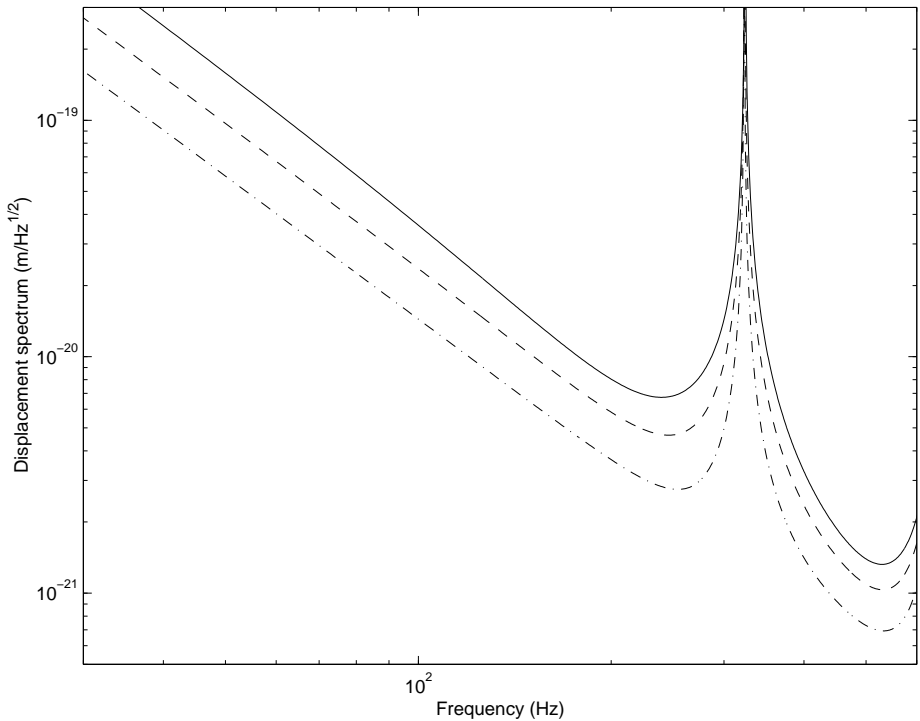}}
\caption{The thermal-noise displacement spectrum 
$\sqrt{x^2 (\omega)}$ for a multi-loop pendulum suspension with
stainless steel wires: $N = 4$ (solid line); 
$N = 8$ (dashed line); $N = 16$ (dash-dot line).}
\end{figure}
%%%%%%%%%%%%%%%%%%%%%%%%%%%%%%%%%%%%%%%%%%%%%%%%%%%%%%%%%%%

\subsection{Comparison between different materials}

We would like to compare the thermal-noise performance of a 
multi-loop suspension for different wire materials. For example,
the tungsten wires examined by Huang and Saulson \cite{HuSa98} 
have rather low breaking stress of 1671 MPa. There exist
tungsten wires with higher breaking stress; for example,
Dawid and Kawamura \cite{DaKa97} experimented with tungsten wires 
for which they measured $\sigma_{{\rm br}} = 2037$ MPa. It would
be interesting to compare between tungsten wires with different
breaking stress but with the same loss function. On the other
hand, the comparison between wires made of tungsten and stainless
steel will clarify how the difference in the loss mechanism
(frequency-independent $\phi_{{\rm w}}$ versus the thermoelastic
damping) affects the thermal-noise spectrum. To this end, we
also would like to consider a situation in which wires made of
stainless steel have all properties as above except for the
losses being dominated by a mechanism with frequency-independent 
$\phi_{{\rm w}}$, instead of the thermoelastic damping.

We consider an eight-loop suspension ($N = 16$), with 
$M = 10.8$ kg, $L = 45$ cm and $\kappa = 0.5$ (like in LIGO),
and examine four possibilities: (a) tungsten wires as considered
in see Sec~3.1, with $\sigma_{{\rm br}} = 1671$ MPa (this 
gives $d \simeq 100$ $\mu$m) and $Q_{{\rm w}} = 4.0 \times 10^3$;
(b) tungsten wires with different breaking stress, 
$\sigma_{{\rm br}} = 2037$ MPa (this gives $d \simeq 91$ $\mu$m) 
and the same quality factor, $Q_{{\rm w}} = 4.0 \times 10^3$;
(c) stainless steel wires as considered in Sec~3.2
($\sigma_{{\rm br}} = 1342$ MPa, $d \simeq 112$ $\mu$m), with
the thermoelastic damping mechanism 
($\beta \simeq 68.6$ s/m$^2$, $\bar{f} \simeq 703$ Hz);
(d) stainless steel wires with the same parameters, but with
a frequency-independent loss function, $Q = 2.0 \times 10^3$
(this value is close to the one given by the thermoelastic 
damping near 120 Hz). The resulting thermal-noise displacement
spectra $\sqrt{x^2 (\omega)}$ are shown in Fig.~3.
One can see that the violin resonances of the stainless steel
wires appear at higher frequencies (due to smaller density).
On the other hand, the tungsten wires exhibit smaller thermal 
fluctuations at the frequency range between 50 and 200 Hz. 
The thermal noise is reduced by using wires with larger breaking 
stress, provided the other parameters remain the same.

%%%%%%%%%%%%%%%%%%%%%%%%%%%%%%%%%%%%%%%%%%%%%%%%%%%%%%%%%%%
\begin{figure}[htbp]
\label{fig:com1}
\epsfxsize=0.72\textwidth
\centerline{\epsffile{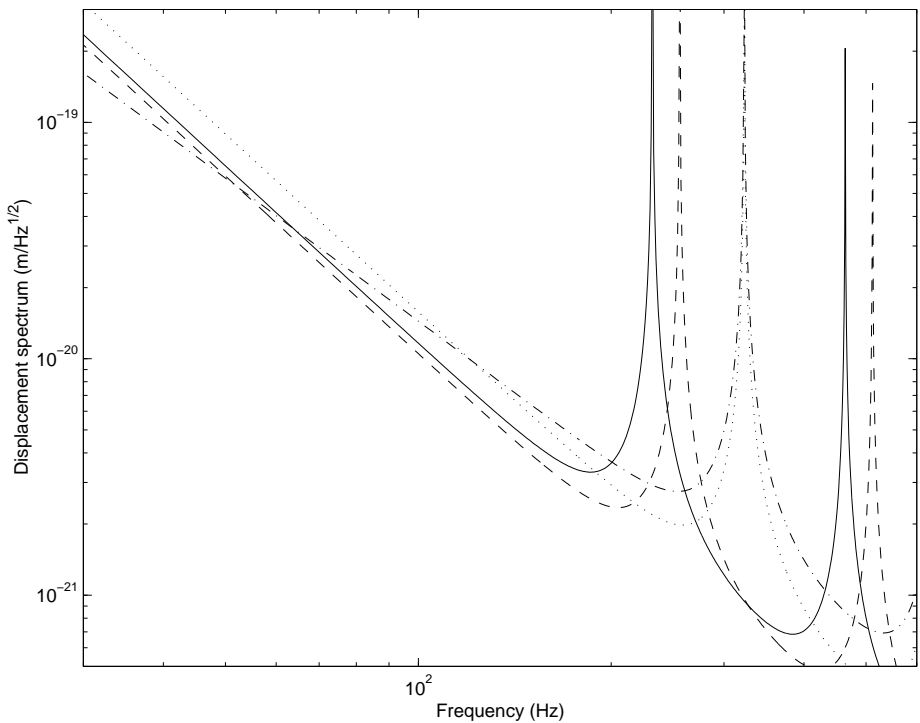}}
\caption{The thermal-noise displacement spectrum 
$\sqrt{x^2 (\omega)}$ for an eight-loop pendulum suspension
($N = 16$): tungsten wires with $\sigma_{{\rm br}} = 1671$ MPa 
and $Q_{{\rm w}} = 4.0 \times 10^3$ (solid line); 
tungsten wires with $\sigma_{{\rm br}} = 2037$ MPa 
and $Q_{{\rm w}} = 4.0 \times 10^3$ (dashed line); 
stainless steel wires with $\sigma_{{\rm br}} = 1342$ MPa 
and thermoelastic damping (dash-dot line);
stainless steel wires with $\sigma_{{\rm br}} = 1342$ MPa
and $Q = 2.0 \times 10^3$ (dotted line).}
\end{figure}
%%%%%%%%%%%%%%%%%%%%%%%%%%%%%%%%%%%%%%%%%%%%%%%%%%%%%%%%%%%

\subsection{Optimization of the pendulum length}

The thermal-noise spectrum depends on the pendulum length $L$.
For frequencies well below the first violin resonance, 
$\omega^2 \ll \omega_1^2$, the pendulum-mode contribution 
dominates and the spectral density $x^2 (\omega)$ is 
proportional to $L^{-2}$. 
However, by increasing $L$, one not only decreases the thermal 
fluctuations due to the pendulum mode, but also brings the 
violin resonances to lower frequencies, as 
$\omega_n \propto L^{-1}$. This effect is illustrated in 
Fig.~4, where the displacement spectrum 
$\sqrt{x^2 (\omega)}$ is shown for an eight-loop suspension
with stainless steel wires of various length. (We take
$M = 10.8$ kg, $\kappa = 0.5$, and stainless steel wires with 
properties listed in Sec.~3.2.) Due to this competition between 
two opposite tendencies, the choice of the pendulum length is 
a delicate matter which depends on where in the spectrum the 
seismic perturbations and the photon shot noise prevail over
the thermal fluctuations and on properties of expected 
gravitational-wave signals.

%%%%%%%%%%%%%%%%%%%%%%%%%%%%%%%%%%%%%%%%%%%%%%%%%%%%%%%%%%%
\begin{figure}[htbp]
\label{fig:ss2}
\epsfxsize=0.72\textwidth
\centerline{\epsffile{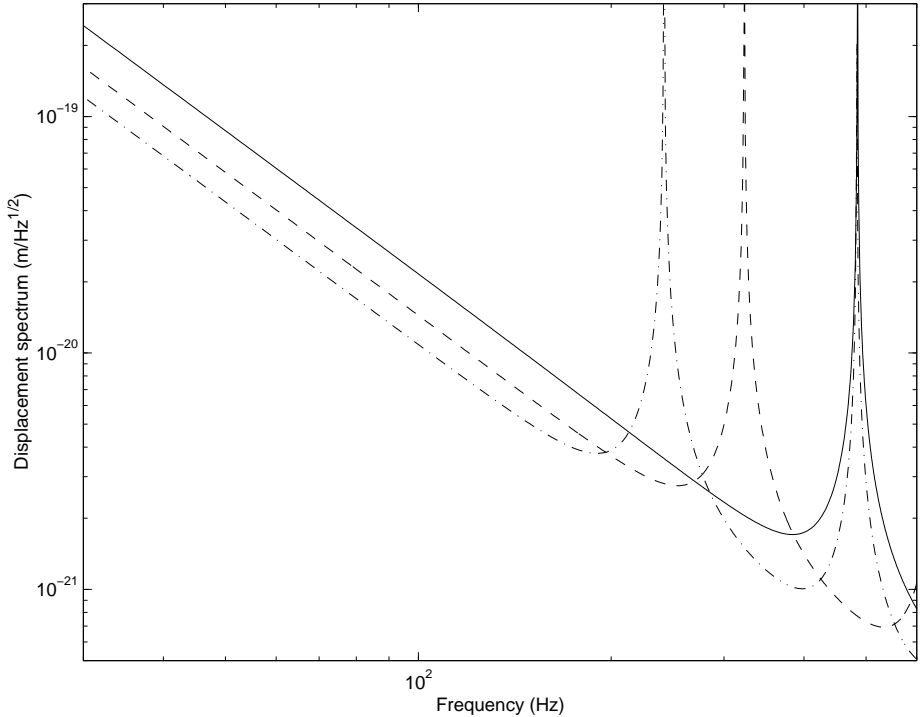}}
\caption{The thermal-noise displacement spectrum 
$\sqrt{x^2 (\omega)}$ for an eight-loop pendulum suspension
($N = 16$) with stainless steel wires of various length: 
$L = 30$ cm (solid line); $L = 45$ cm (dashed line); 
$L = 60$ cm (dash-dot line).}
\end{figure}
%%%%%%%%%%%%%%%%%%%%%%%%%%%%%%%%%%%%%%%%%%%%%%%%%%%%%%%%%%%

\section{Discussion}

Our analysis brings to an observation that the thermal noise in 
pendulum suspensions can be significantly reduced by using 
multi-loop configurations with a large number of wires. However,
before implementing this conclusion one should consider a number
of issues. First, our analysis is valid only if the losses are 
dominated by the internal friction in the pendulum wires and all
other sources of dissipation are made negligible by careful
experimental design. However, as was shown recently by Huang and 
Saulson \cite{HuSa98}, the sliding friction in the suspension 
clamps is often important as well. If this is the case, a large 
number of suspension loops will only sever the dissipation and 
thereby increase the thermal fluctuations. Therefore, if one 
wants to use multi-loop suspensions, a special care should be 
paid to the design of clamps. Another technical problem is to 
make a suspension in which all the loops will be equally loaded. 
One more issue which should be carefully studied is the effect
which may have a large number of suspension wires on the 
internal resonances of the suspended test mass.

\section*{Acknowledgments}

This work would not be possible without great help by Malik 
Rakhmanov. I thank him for long hours of illuminating discussions 
and for encouraging me to enter the realm of thermal
noise and anelasticity. 
I am also grateful to Peter Saulson and Gregg Harry
for sending me their data on properties of wire materials.
Financial support from the Lester Deutsch Fund in the form of
a postdoctoral fellowship is gratefully acknowledged.
The LIGO Project is supported by the National Science Foundation 
under the cooperative agreement PHY-9210038.

\end{document}